\documentclass[aps,pre,amsmath,amssymb,
 preprint,superscriptaddress
]{revtex4-1}
\allowdisplaybreaks
\usepackage{graphicx}
\usepackage{dcolumn}

\usepackage{mathptmx}
\usepackage{siunitx,color,bm,epsfig,mathtools,MnSymbol,chemfig}

\newcommand{\kt}{k_{\mathrm{B}}T}

\begin{document}

\title[Assembly of heteropolymers via { a network of}  reaction coordinates]{{Assembly of heteropolymers via a network of reaction coordinates}}

\author{D. Chiuchi\`u}
 \affiliation{Biological complexity unit, Okinawa Institute for Science and Technology, 1919-1 Tancha, Onna, Kunigami-gun, Okinawa 904-0412, Japan}
\author{James Ferrare}
\affiliation{Biological complexity unit, Okinawa Institute for Science and Technology, 1919-1 Tancha, Onna, Kunigami-gun, Okinawa 904-0412, Japan}%
 \affiliation{Tulane University, 6823 St Charles Ave, New Orleans, LA 70118, USA}
\author{S. Pigolotti}%
 \email{simone.pigolotti@oist.jp}
\affiliation{Biological complexity unit, Okinawa Institute for Science and Technology, 1919-1 Tancha, Onna, Kunigami-gun, Okinawa 904-0412, Japan}%
\date{\today}

\begin{abstract}
In biochemistry, heteropolymers encoding biological information are assembled out of equilibrium by sequentially incorporating available monomers found in the environment. Current models of polymerization treat monomer incorporation as a sequence of discrete chemical reactions between intermediate meta-stable states. In this paper, we use ideas from reaction rate theory and describe non-equilibrium assembly of a heteropolymer via a continuous reaction coordinate. Our approach allows to estimate the copy error and incorporation speed from the Gibbs free energy landscape of the process. We apply our theory to several examples, from a simple reaction characterized by a free energy barrier to more complex cases incorporating error correction mechanisms such as kinetic proofreading.
\end{abstract}

\maketitle

\section{Introduction}
DNA, RNA, and proteins are the building blocks of all living systems. These heteropolymers are assembled to match a template; only a very small number of mismatches with the template is tolerable for maintaining biological information and for correct functioning of cells. However, the binding energies of different monomers usually differ by only a few $k_BT$, where $k_B$ is the Boltzmann constant and $T$ the temperature. This means that, at physiological temperature, mismatches can not be completely suppressed \cite{Reynolds2010}. 

\begin{figure*}
\includegraphics[width=\linewidth]{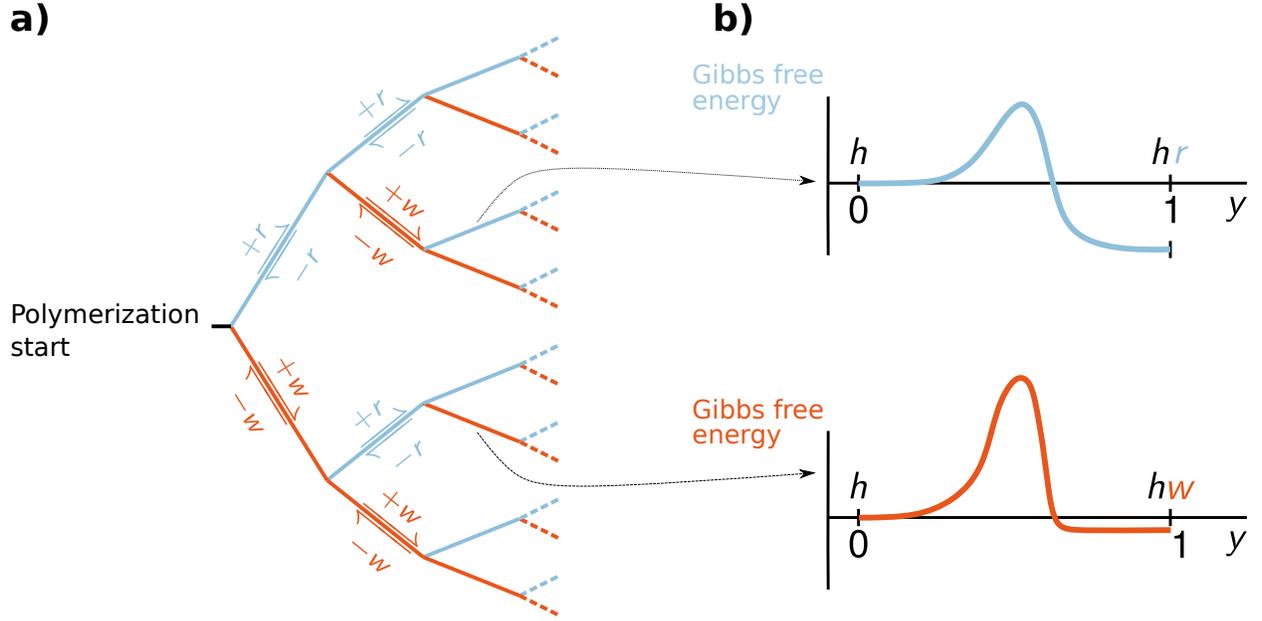}
\caption{Synthesis of heteropolymers. a) Network of incorporation and removal reactions to synthesize a heteropolymer. Each edge in the network represent the chemical reaction $h \rightleftharpoons  hx$ where $h$ is the heteropolymer produced so far and $hx$ is the same heteropolymer with addition of monomer $x\in\{r,w\}$ at the tip. Three reactions compete at the nodes of the network: removal of the last incorporated monomer, incorporation of a $r$ monomer, and incorporation of a $w$ monomer. b) Reaction coordinate description of the incorporation and removal reactions. The initial and final points of the free energy landscapes correspond to the reactants and products of the incorporation and removal reactions, respectively.}\label{fig:tree}
\end{figure*}

Our aim is to describe the chemical processes responsible for these errors. Specifically, we consider sequential assembly of heteropolymers where each incorporated monomer can be a right ($r$) or a wrong ($w$) match with a template. These two { different outcomes }can be { represented as competing} chemical { reactions}
\begin{equation}\label{eq:incorporation_reaction}
\schemestart
 h
 \arrow(@c1--){<=>[+w][-w]}[-30] hw 
 \arrow(@c1--){<=>[+r][-r]}[30] hr
\schemestop
\end{equation}
where $h$ is the heteropolymer produced so far, and $hr$/$hw$ are the same heteropolymer with an addition of a $r$/$w$ monomer at the tip, respectively. Each monomer incorporation {is} iteratively followed by a new one, so that the whole polymerization process {is} described by the tree-shaped network of chemical reactions  \cite{bennett1979dissipation, Pigolotti2016} in Fig. \ref{fig:tree}a. 

To achieve accurate and fast assembly, the reactions { in Eq.\eqref{eq:incorporation_reaction} involve several intermediate steps, }such as initial monomer discrimination \cite{doi:10.1146/annurev.biochem.69.1.497}, kinetic proofreading, \cite{Hopfield4135, NINIO1975587, doi:10.1146/annurev.biochem.69.1.497}, and mismatch repair \cite{ODonnell01072013, doi:10.1146/annurev-genet-112414-054722}. { In general, each one of these error-correction mechanisms contribute simultaneously to p}olymerization accuracy, speed \cite{doi:10.1146/annurev.biochem.70.1.415, andrieux2008nonequilibrium, JOHANSSON2008141, doi:10.1063/1.4890821, Banerjee5183, PhysRevX.5.041039, savir2013ribosome, PhysRevLett.110.188101, murugan2012speed}, and  energetic cost \cite{cady2009open, KRAMERS1940284, 1742-5468-2015-6-P06001, PhysRevX.5.041039, Wagoner5902}. 

{ Two approaches can provide insight into the  error-correction mechanisms underlying heteropolymer assembly. The first approach is to measure their kinetic rates under different experimental conditions \cite{doi:10.1146/annurev.biochem.70.1.415}. The second} approach is to simulate heteropolymer assembly using molecular dynamics \cite{BOCK201827}. From the molecular dynamics, one can project the numerous degrees of freedom into a \mbox{1-dimensional} collective variable called reaction coordinate \cite{doi:10.1002/wcms.1276}. The reaction coordinate simplifies a chemical process into a one-dimensional random motion \cite{doi:10.1002/wcms.1276, doi:10.1063/1.471317, PhysRevLett.96.228104}. The parameters of this random motion depend on the underlying reactants dynamics  \cite{PhysRev.124.983, doi:10.1021/jp403043a, doi:10.1063/1.4890367} and on the projection technique \cite{doi:10.1063/1.4890367, doi:10.1002/wcms.1276, doi:10.1021/jp060039b}.

While successful in describing  protein folding { \cite{doi:10.1063/1.471317, PhysRevLett.96.228104, PhysRevLett.79.317}} { and in modeling reaction rates \cite{KRAMERS1940284, PhysRevLett.79.317}}, approaches based on reaction coordinates found little use in studies of polymerization speed and accuracy. In principle, both reactions in Eq. \eqref{eq:incorporation_reaction} can be described by means of a reaction coordinate (Fig. \ref{fig:tree}b). However, { to study the complete polymerization process we need to join the reaction coordinates characterizing each branch in Fig. \ref{fig:tree}a. Mathematically, this amounts to impose appropriate boundary conditions at the nodes of the reaction network.}

In this paper, we develop a model of heteropolymer assembly { based on reaction coordinates}, and  use it to compute the accuracy and speed of polymerization in different conditions. The paper is organized as follows. In Section \ref{sec:model}, we introduce our model. From the reaction coordinate, we derive effective incorporation and removal probabilities of right and wrong monomers. In Section \ref{sec:results}, we compute the accuracy and speed of a general heteropolymer assembly. In Section \ref{sec:examples} we consider examples characterized by different Gibbs free energy landscapes. In Section \ref{sec:kinetic_proofreading}, we generalize our results to a case where the reaction leading to monomer incorporation is complemented by kinetic proofreading. Section \ref{sec:conclusions} is devoted to conclusions and perspectives.

\section{Model}\label{sec:model}

{ We define our model of heteropolymer assembly with reaction coordinates through the following steps. We first introduce the reaction coordinate and the free energy landscape in each chemical reaction of the polymerization network. We then  study the dynamics of the reaction coordinate dynamics and its boundary conditions at the nodes of the network. Finally, we compute the probabilities to incorporate/remove one monomer along each reaction coordinate.}

\subsection{Reaction coordinate and Gibbs free energy of the heteropolymer}\label{sec:rcgibbs}

We { introduce} the continuous reaction coordinate $y$ along each edge of the polymerization network, Fig. \ref{fig:tree}a. Without loss of generality,  we choose the units of the reaction coordinate  so that $y\in[0,1]$, where $y=0$ and $y=1$ correspond to $h$ and $hx$ respectively, i.e. to the states  before and after monomer incorporation, see Figure \ref{fig:tree}.b.

Each point along this reaction coordinate is characterized by a Gibbs free energy $G^{hx}(y)$ (from now on simply "free energy"). Such free energy depends on the previously incorporated sequence of monomers ($h$), on the candidate monomer to be incorporated ($x$) and on the stage of the incorporation process, i.e. the value of $y$. { Implicitly, $G^{hx}(y)$ also depends on the reactant and product concentrations. }

\begin{figure}
    \includegraphics[width=\linewidth]{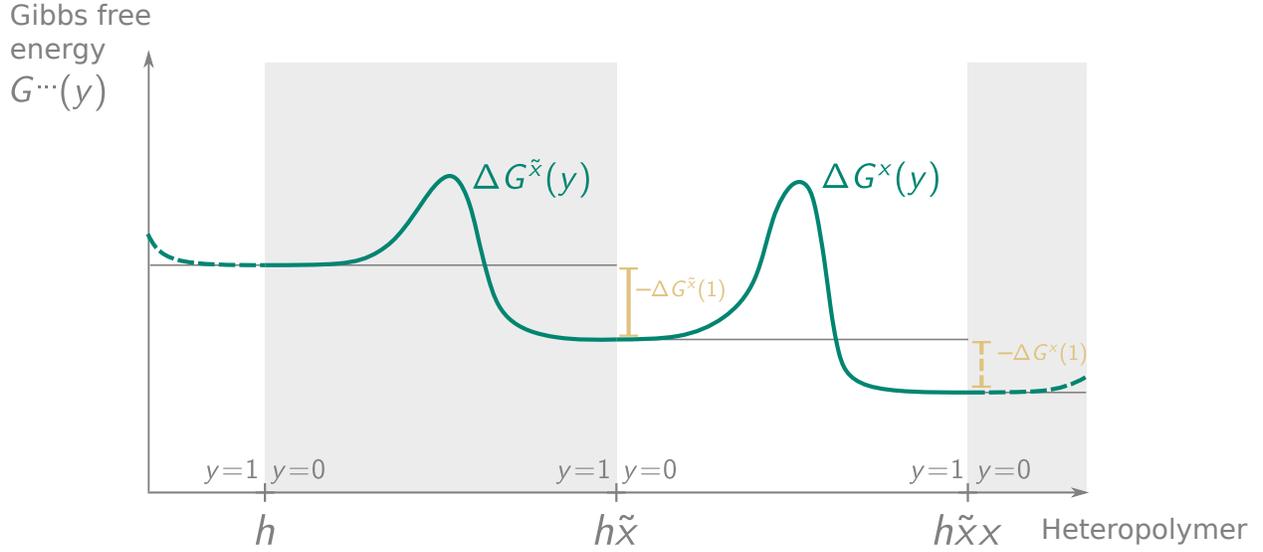}
    \caption{Free energy $G^{\ldots}(y)$ of the heteropolymer depends on the reaction coordinate $y$, and the sequence of incorporated monomers. The superscript of the free energy indicates the stage of the polymerization process (either $h$, $h\tilde{x}$, or $h\tilde{x}x$ in this case). { The functions $\Delta G^{\tilde{x}}(y)$ and $\Delta G^{x}(y)$ are the free energy increments along the reaction coordinate to incorporate $\tilde {x}$ and $x$ respectively. The total binding free energies for monomer $\tilde{x}$ and $x$ are $-\Delta G^{\tilde{x}}(1)$ and $-\Delta G^x(1)$, respectively.}}
    \label{fig:continuous_energy}
\end{figure}

{ 

We introduce the free energy increments from the beginning of each incorporation reaction
\begin{equation}\label{eq:Gibbs_increment}
    \Delta G^x(y)=G^{hx}(y)-G^{hx}(0),
\end{equation}
see Fig. \ref{fig:continuous_energy}. The free energy increments depend on the candidate monomer $x$ but not on the whole history of incorporated monomers $h$. With this notation, the (absolute) binding free energy of monomer $x$ is equal to $-\Delta G^x(1)$.

The free energy must be a continuous function of $y$, and must also vary continuously when crossing the nodes of the network in Fig. \ref{fig:tree}. This means that we can decompose the free energy at an arbitrary stage of the polymerization process as
\begin{equation}\label{eq:free_energy_quasi_continuity}
\begin{aligned}
    G^{hx}(y)=&G^{hx}(0)+\Delta G^{x}(y)\\
             =&G^{h}(1)+\Delta G^{x}(y)\\
             =&\left(\sum_{i\in h} \Delta G^{i}(1) \right) + \Delta G^{x}(y).
\end{aligned}
\end{equation}
}

\subsection{Stochastic dynamics of the reaction coordinate { and boundary conditions}}

Because of thermal fluctuations, the reaction coordinate $y$ evolves according to a Langevin equation
\begin{equation}\label{eq:Langevin_dynamics}
\frac{\mathrm{d}y}{\mathrm{d} t}=-\mu\frac{\mathrm{d}}{\mathrm{d} y} G^{hx}(y) +\sqrt{2 D}~\xi(t)
\end{equation}
where $\mu$ is a mobility, $D$ is a diffusion coefficient, and $\xi(t)$ is white noise with $\langle \xi(t)\rangle=0$ and $\langle \xi(t)\xi(t')\rangle=\delta(t-t')$ \cite{gardiner2009stochastic}. We assume that $D$ satisfies the Einstein relation \mbox{$D=k_BT \mu$} with temperature $T$ and Boltzmann constant $k_B$. { We also assume that $D$, $\mu$, and $T$ are constant.} When the reaction coordinate reaches the boundaries, either $y=0$ or $y=1$, a new incorporation/removal reaction is commenced.  

Equation \eqref{eq:Langevin_dynamics} needs to be complemented by rules to specify which reaction initiates at the nodes of the { reaction} network. To this aim,  we consider two intermediate values of the reaction coordinate: $y=\epsilon$ and $y=1-\epsilon$ with $\epsilon\ll1$. Using these values we coarse-grain the evolution of the reaction coordinate $y$ as
\begin{equation}\label{eq:four_state_dynamics}
(y=0) \xrightleftharpoons[\pi_{\epsilon,0}^x]{\pi_{0,\epsilon}^x} (y=\epsilon) \xrightleftharpoons[\pi_{1-\epsilon,\epsilon}^x]{\pi_{\epsilon,1-\epsilon}^x} (y=1-\epsilon) \xrightleftharpoons[\pi_{1,1-\epsilon}^x]{\pi_{1-\epsilon,1}^x} (y=1)
\end{equation}
{ where t}he quantities $\pi_{\tilde{y},y}^{x}$ are first-passage probability  from $y$ to $\tilde{y}$. For example, $\pi_{1-\epsilon,\epsilon}^x$ is the probability that the reaction coordinate reaches $y=1-\epsilon$ from $y=\epsilon$   without having reached $y=0$ before.

{The representation in Eq.\eqref{eq:four_state_dynamics}} separates the dynamics in proximity of the nodes of Fig. \ref{fig:tree}.b, from the dynamics in the interval $[\epsilon,1-\epsilon]$. Thanks to this separation, we {  use detailed balance, probability conservation close to the nodes, and the continuity of $G^{hx}(y)$ to } compute the first-passage probabilities { $\pi_{\epsilon,0}^x$ and $\pi_{1-\epsilon,1}^{x}$ (see Appendix \ref{sec:nodes}). This procedure results in
\begin{equation}\label{eq:pis_initiation energy}
\pi_{\epsilon,0}^x= \pi_{1-\epsilon,1}^{x} = \frac{1}{3}+\mathcal{O}(\epsilon).
\end{equation}
We compute the first-passage probabilities in the interior applying standard techniques  \cite{gardiner2009stochastic, doi:10.1002/9781119165156.ch5} to the Fokker-Planck equation associated to Eq. \eqref{eq:Langevin_dynamics}. We obtain
\begin{eqnarray}\label{eq:pis_first_passage}
\pi_{0,\epsilon}^{x} &=&\frac{\int_{\epsilon}^{1-\epsilon}\psi^{x}(y)\mathrm{d}y}{\int_{0}^{1-\epsilon}\psi^{x}(y)\mathrm{d}y}\nonumber\\
\pi_{\epsilon,1-\epsilon}^{x} &=&\frac{\int_{1-\epsilon}^{1}\psi^{x}(y)\mathrm{d}y}{\int_{\epsilon}^{1}\psi^{x}(y)\mathrm{d}y}\nonumber\\
\pi_{1-\epsilon,\epsilon}^{x} &=&\frac{\int_{0}^{\epsilon}\psi^{x}(y)\mathrm{d}y}{\int_{0}^{1-\epsilon}\psi^{x}(y)\mathrm{d}y}\nonumber\\
\pi_{1,1-\epsilon}^{x} &=&\frac{\int_{\epsilon}^{1-\epsilon}\psi^{x}(y)\mathrm{d}y}{\int_{\epsilon}^{1}\psi^{x}(y)\mathrm{d}y}
\end{eqnarray}
with 
\begin{equation}\label{eq:psi}
\begin{aligned}
\psi^{x}(y)=&\exp\left[\int_0^y\frac{\mu}{D}\frac{\partial \Delta G^x(z)}{\partial z} \mathrm{d}z\right]\\ 
=&\exp\left[\frac{\Delta G^{x}(y)}{\kt}\right]
\end{aligned}
\end{equation}
where the last equality follows from the relation $D = k_B T \mu$.

\subsection{Effective probabilities of monomer incorporation/rejection}
From the probabilities $\pi_{\tilde{y},y}^{x}$, we now compute the }effective probabilities { $p_\rightarrow^x$ and $p_\leftarrow^x$  to incorporate and reject monomer $x$ along each edge of the reaction network in Figure \ref{fig:tree}. To this end, we assume that the coarse grained dynamics in Eq.\eqref{eq:four_state_dynamics} is at steady state. We then use adiabatic elimination \cite{doi:10.1063/1.2907242} to obtain (see  Appendix \ref{sec:effective_probabilities})}
\begin{subequations}\label{eq:p_reaction_coordinate}
\begin{align}
p_\rightarrow^{x}=&{\frac{\epsilon}{3}}\,\frac{1}{\int_{0}^{1}\exp[\frac{\Delta G^{x}(y)}{\kt}]\,\mathrm{d}y}+\mathcal{O}(\epsilon)\label{eq:incorporation_rate}\\
p_\leftarrow^{x}=&{\frac{\epsilon}{3}}\,\frac{e^{\frac{\Delta G^{x}(1)}{\kt}}}{\int_{0}^{1}\exp\left[\frac{\Delta G^{x}(y)}{\kt}\right]\,\mathrm{d}y}+\mathcal{O}(\epsilon)
\end{align}
\end{subequations}

Equations \eqref{eq:p_reaction_coordinate} relate the free energy landscapes $G^x(y)$ and the incorporation/removal probabilities of the polymerization process. They are consistent with the detailed balance condition
\begin{equation}\label{eq:detailed_balance}
\frac{p^{x}_{\rightarrow}}{p_\leftarrow^{x}}=\exp\left[-\frac{ \Delta G^x(1)}{\kt}\right].
\end{equation}
which connects the ratios of forward and backward probabilities to the binding free energy {$-\Delta G^x(1)$}, see Fig.  \ref{fig:continuous_energy}.

\section{Results}\label{sec:results}

We now  address the accuracy and speed of a polymerization process in the reaction coordinate framework. We consider a copy polymer made up of a number $N^r$ of right monomers and $N^w$ of wrong monomers with $N=N^r+N^w$. For large $N$, we define the error rate 

\begin{equation}\label{eq:eta_def}
\eta=\lim_{N\rightarrow\infty}\frac{N^w}{N}.
\end{equation}
 To compute $\eta$ from the incorporation and removal probabilities $p_\rightarrow^x$ and $p_\leftarrow^x$, we first recast Eq.\eqref{eq:eta_def} into the implicit equation
\begin{equation}\label{eq:proto_eta_equation}
\frac{\eta}{1-\eta}=\lim_{N\rightarrow\infty}\frac{N^w}{N^r}= \lim_{n\rightarrow\infty}\frac{n^w_\rightarrow-n^w_\leftarrow}{n^r_\rightarrow-n^r_\leftarrow}
\end{equation}
where we have introduced the numbers $n_\rightarrow^r$, $n_\leftarrow^r$, $n_\rightarrow^w$ and $n_\leftarrow^w$ of $r$ and $w$ incorporation and removal reactions which have occurred in the process, and $n$ is the total number of observed chemical reactions. For large $n$ we have 
\begin{subequations}\label{eq:count_reactions}
\begin{align}
n_\rightarrow^r\sim~& n\, c\, p_\rightarrow^r\\
n_\leftarrow^r\sim~& n\, c\,(1-\eta)\, p_\leftarrow^r\label{eq:Nleftr}\\
n_\rightarrow^w\sim~& n\, c\,p_\rightarrow^w\\
n_\leftarrow^w\sim~& n\, c\,\eta\, p_\leftarrow^w\label{eq:Nleftw}
\end{align}
\end{subequations}
where $c=[p_\rightarrow^r+(1-\eta) p_\leftarrow^r+p_\rightarrow^w+\eta p_\leftarrow^w]^{-1}$ is a normalization constant so that $n=n_\rightarrow^r + n_\leftarrow^r +n_\rightarrow^w + n_\leftarrow^w=n$. Substituting Eqs. \eqref{eq:count_reactions} into Eq. \eqref{eq:proto_eta_equation} gives
\begin{equation}\label{eq:eta_equation}
\frac{\eta}{1-\eta}=\frac{p^w_\rightarrow-\eta p^w_\leftarrow}{p^r_\rightarrow-(1-\eta)p^r_\leftarrow}
\end{equation}

Equation \eqref{eq:eta_equation} is a general "self-consistency" relation for the error rate that holds also for discrete models of polymerization \cite{PhysRevX.5.041039, Pigolotti2016, bennett1979dissipation}. 
In our case, we substitute Eqs.~\eqref{eq:p_reaction_coordinate} in Eq.~\eqref{eq:eta_equation} and take the limit $\epsilon\to 0$, obtaining
\begin{equation}\label{eq:eta_single_step}
\begin{aligned}
\frac{\eta}{1-\eta}=&\left(\frac{\eta-\exp\left[{-\frac{ \Delta G^w(1)}{\kt}}\right]}{1-\eta-\exp\left[{-\frac{ \Delta G^r(1)}{\kt}}\right]}\right)\ \left( \exp\left[{\tfrac{\Delta G^w(1)-\Delta G^{r}(1)}{\kt}}\right]\ \frac{\int_{0}^1 \exp\left[\frac{\Delta G^r(y)}{\kt}\right]\,\mathrm{d}y}{\int_{0}^1 \exp\left[\frac{\Delta G^w(y)}{\kt}\right]\,\mathrm{d}y}\right).
\end{aligned}
\end{equation}
{ Solving Eq. \ref{eq:eta_single_step} for $\eta$ yields an explicit expression of the error rate from the energy potentials.}
 
Equation \ref{eq:eta_single_step} { identifies different regimes of error correction.  To identify a first regime, }we observe that
\begin{equation}
    \eta=\frac{1}{1+\exp\left[\frac{ \Delta G^w(1)-\Delta G^r(1)}{\kt}\right]} \quad \mbox{if }\quad  \exp\left[{\tfrac{\Delta G^w(1)-\Delta G^{r}(1)}{\kt}}\right]\ \frac{\int_{0}^1 \exp\left[\frac{\Delta G^r(y)}{\kt}\right]\,\mathrm{d}y}{\int_{0}^1 \exp\left[\frac{\Delta G^w(y)}{\kt}\right]\,\mathrm{d}y}=1, \label{eq:energetic}
\end{equation}
In the regime where Eq. \eqref{eq:energetic} holds, the error depends only on the binding free energy difference $\Delta G^w(1)-\Delta G^r(1)$. This regime is called energetic discrimination regime in the literature \cite{PhysRevLett.110.188101, Pigolotti2016}.  Systems near equilibrium operates in this regime because the Boltzmann factors of the binding free energies determine, via detailed balance, the probabilities  to incorporate different monomers.

{ To identify a second error-correction regime in Eq. \eqref{eq:eta_single_step}, w}e consider the case where $\Delta G^r(y)$ and $\Delta G^w(y)$ are characterized by energy barriers with heights $\delta^r$ and $\delta^w$ respectively (see Figure \ref{fig:tree}.b and Kramers \cite{KRAMERS1940284}). When such barriers are large, we can approximate the integrals in Eq. \eqref{eq:eta_single_step} by using the Laplace method \cite{bender1978advanced}
\begin{equation}\label{eq:integrals_saddle}
\frac{\int_{0}^1 \exp\left[\frac{\Delta G^r(y)}{\kt}\right]\,\mathrm{d}y}{\int_{0}^1 \exp\left[\frac{\Delta G^w(y)}{\kt}\right]\,\mathrm{d}y}\approx\exp\left[\frac{\delta^r-\delta^w}{\kt}\right]\sqrt{\frac{\Sigma^w}{\Sigma^r}}
\end{equation}
where  $\Sigma^r$ and $\Sigma^w$ are the curvatures of $\Delta G^r(y)$ and $\Delta G^w(y)$ at their maxima, respectively. { Equation \ref{eq:integrals_saddle} implies that activation barriers  suppress the polymerization error via the second term in round brackets in Eq. \eqref{eq:eta_single_step}. The regime where this suppression occurs is the kinetic discrimination regime \cite{PhysRevLett.110.188101, Pigolotti2016}. The first factor on the right-hand side of Eq. \eqref{eq:integrals_saddle} represents the contribution of a difference in activation energy barrier $\delta^r-\delta^w$ to accuracy. } This effect is also present in models based on discrete-step reactions \cite{bennett1979dissipation,cady2009open,PhysRevLett.110.188101, Pigolotti2016}. The factor $\sqrt{\Sigma^w/\Sigma^r}$ is a correction to activation energies based on the width of the activation barriers. { This factor permits kinetic discrimination at equal barrier heights}, provided that the  barrier for right monomers is significantly more narrow than for wrong monomers.

We estimate the average polymerization speed
using a similar argument to that leading to Eq. \eqref{eq:eta_single_step}.
For large number $N$ of incorporated monomers, the average speed is equal to $N$ divided the total time $\mathcal{T}$ needed to assemble the polymer
\begin{equation}\label{eq:proto_v_definition}
\begin{aligned}
    v=&\lim_{N\to\infty} \frac{N}{\mathcal{T}}\\
    =& \lim_{N\to\infty} \frac{\left(n_\rightarrow^r-n_\leftarrow^r\right) + \left(n_\rightarrow^w-n_\leftarrow^w\right)}{\mathcal{T}}
\end{aligned}
\end{equation}
where we expressed $N$ in terms of the number of incorporation/removal reactions. For large $N$ we can approximate the polymerization time as
\begin{equation}\label{eq:tau_approx}
    \mathcal{T} \sim n  \langle \tau \rangle
\end{equation}
where $\langle \tau \rangle$ is the average time it takes to either incorporate or remove a monomer. 
Substituting Eqs. \eqref{eq:count_reactions} and \eqref{eq:tau_approx} into Eq. \eqref{eq:proto_v_definition} gives the estimate for the polymerization speed
\begin{equation}\label{eq:v_definition}
v=\frac{c[p_\rightarrow^r-(1-\eta)p_\leftarrow^r+ p_\rightarrow^w-\eta p_\leftarrow^w]}{\langle\tau\rangle}.
\end{equation}

{ The numerator of Eq. \eqref{eq:proto_v_definition}} is the probability of an incorporation minus the probability of a removal, { while} the denominator provides the timescale of these events. In practice, calculating $\langle \tau\rangle$ is not straightforward since one has to take into account contributions from incorporation attempts that are not finalized. In { Appendix \ref{sec:derive_velocity}}, we provide a more formal derivation of Eq. \eqref{eq:v_definition}, together with an explicit expression for $\langle \tau\rangle$.

\section{Examples}\label{sec:examples}
To address the validity and practical implications of Eqs. \eqref{eq:eta_single_step} and \eqref{eq:v_definition} we consider two examples of potentials $\Delta G^{r}(y)$ and $\Delta G^w(y)$. { In both cases, we work in dimensionless units by fixing $D=1$, $\kt=1$,  and $\mu=1$.}

\subsection{Linear potential}  

As first example we consider linear free energy landscapes
\begin{subequations}\label{eq:linear_potentials}
\begin{align}
\Delta G^{r}(y)=-m_r\, y\\
\Delta G^{w}(y)=-m_w\, y .
\end{align}
\end{subequations}

\begin{figure*}
\includegraphics[width=\linewidth]{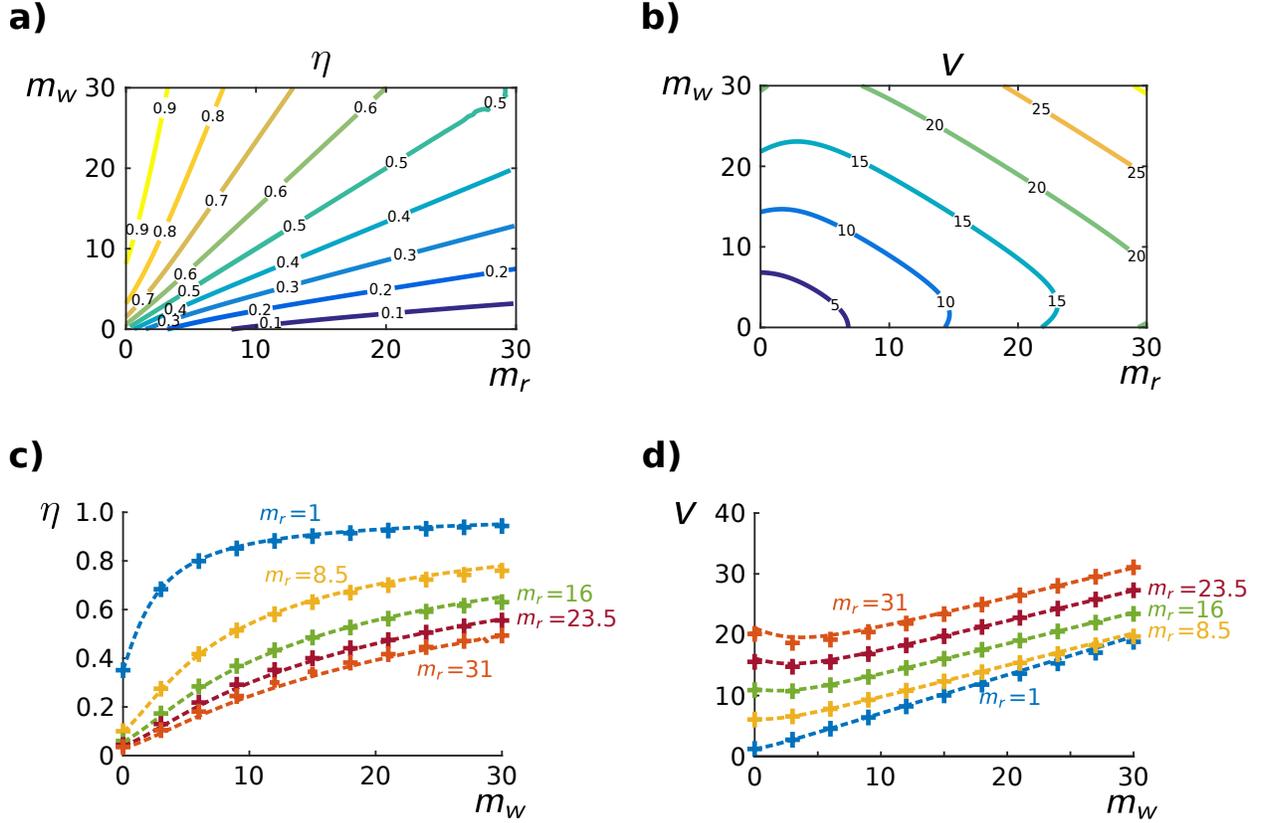}
\caption{Error rate $\eta$ and velocity $v$ for linear free energy landscapes . (Top) surface plots of $\eta$ and $v$ from Eqs.  \eqref{eq:eta_single_step} and \eqref{eq:v_definition} when $G^r(y)=-m_r y$ and $G^w(y)=-m_w y$ as a function of the irreversibility parameters $m_r$ and $m_w$. (Bottom) Error $\eta$ and  speed $v$ as a function of $m_w$ for different values of $m_r$. Crosses represent the average $\eta$ and $v$ values measured from $700$ numerical simulations of the stochastic incorporation process. The Langevin dynamics of Eq. \eqref{eq:Langevin_dynamics} was integrated with the  Euler-Maruyama  scheme \cite{kloeden2011numerical}. \label{fig:numerical_simulations}}
\end{figure*}

Despite their simplicity, the potentials in Eq. \eqref{eq:linear_potentials} are useful to understand the physics of the process. Upon increasing the slopes $m_r$ and $m_w$, polymerization becomes increasingly irreversible.  Substituting the potentials Eq. \eqref{eq:linear_potentials} into the expression for the error,  Eq. \eqref{eq:eta_single_step} and performing the integrals we obtain
\begin{equation}\label{eq:error_linear}
    \frac{\eta}{1-\eta}=\frac{m_w(1-e^{-m_r})[1-e^{-m_w}\eta]}
    {m_r(1-e^{-m_w})[1+e^{-m_r}(1-\eta)]},
\end{equation}
which implies
\begin{equation}\label{eq:error_linear2}
    \eta =\frac{m_w}{m_r+m_w}\qquad \mbox{for}\qquad m_r,m_w\gg 1.
\end{equation}
The exact solution of Eq. \eqref{eq:error_linear} shows that the error is approximately a function of $m_w/m_r$ when $m_r$, $m_w$ are large, as predicted by Eq. \eqref{eq:error_linear2}, Fig. \ref{fig:numerical_simulations}a. We compared the predictions from Eqs. \eqref{eq:error_linear} and \eqref{eq:v_definition} with numerical simulations of the incorporation process from Eq.\eqref{eq:Langevin_dynamics}. Our theory yields reliable predictions for a broad range of  parameters, Fig. \ref{fig:numerical_simulations}c and \ref{fig:numerical_simulations}d.

\begin{figure}
\includegraphics[width=0.5\linewidth]{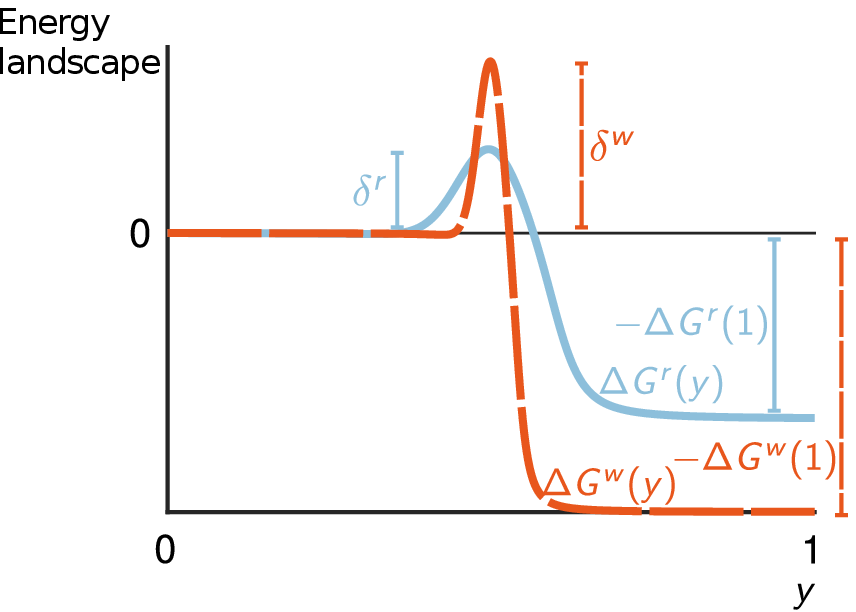}
\caption{Free energy potentials with a barrier for $r$ and $w$ monomers from Eq. \eqref{eq:barrier_potential} for $r$ and $w$ monomers. We chose the parameters so that the free energy landscapes for $r$ and $w$ monomers have different binding energies ($-\Delta G^r(1)$ and $-\Delta G^w(1)$), different barrier heights ($\delta^r$ and $\delta^w$), and different barrier widths ($\sigma^r$ and $\sigma^w$). }
\label{fig:barrier_potential_fig}
\end{figure}

\subsection{Potential with an activation barrier} 

As a second example we consider the potential
\begin{equation}\label{eq:barrier_potential}
\begin{aligned}
\Delta G^x(y)=&a_x \left(e^{-\frac{\left(y-\frac{1}{2}\right)^2}{2c_x^2}}-e^{-\frac{1}{8 c_x^2}}\right)+\frac{b_x}{2}\left( \frac{2 c_x+\frac{1}{2}-y}{\sqrt{\left(2 c_x+\frac{1}{2}-y\right)^2+c_x^2}}-\frac{2 c_x+\frac{1}{2}}{\sqrt{\left(2 c_x+\frac{1}{2}\right)^2+c_x^2}} \right)
\end{aligned}
\end{equation}
where $a_x$, $b_x$ and $c_x$ are monomer-dependent parameters that control the shape of the free energy potentials.  { Key features of the potential of Eq. \eqref{eq:barrier_potential} are the binding energy $-\Delta G^x(1)$, the height of the activation barrier $\delta^x$ and its width $\sigma^x=4c_x$, Fig. \ref{fig:barrier_potential_fig}.}

\begin{figure*}
\includegraphics[width=\linewidth]{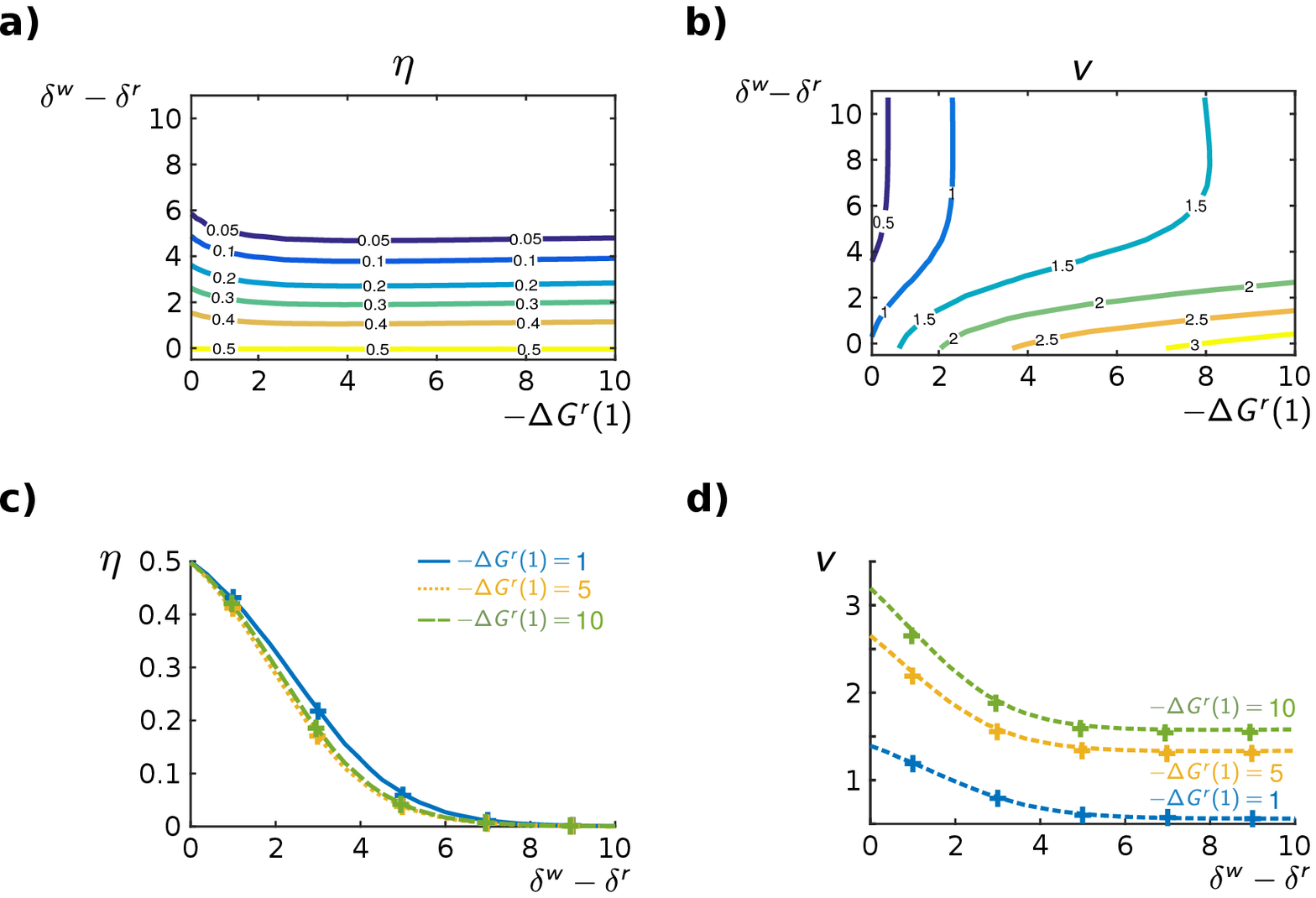}
\caption{Equations \eqref{eq:eta_single_step} and \eqref{eq:v_definition} predict $\eta$ and $v$ in a kinetic discrimination regime. (Top) contour plots of $\eta$ and $v$ from Eqs.  \eqref{eq:eta_single_step} and \eqref{eq:v_definition} as a function of the activation energy difference $\delta^w-\delta^r$ and the binding energy $-\Delta G^r(1)$. In this example, we chose $a_r=1$, $b_r=b_w$ and $c_r=c_w=1/20$ to ensure a kinetic discrimination regime where $\Delta G^r(1)=\Delta G^w(1)$. Large values of $-\Delta G^r(1)$ correspond to highly irreversible processes. (Bottom) Plots of $\eta$ and $v$ versus the activation energy difference $\delta^w-\delta^r$ at fixed values of $\Delta G^r(1)$. Crosses corresponds to the average values of $\eta$ and $v$  measured from $300$ stochastic simulations of the incorporation process with Eq.\eqref{eq:Langevin_dynamics}. The Langevin dynamics was simulated with a weak 2.0 Runge-Kutta stochastic scheme \cite{kloeden2011numerical}. \label{fig:numerical_simulations_barrier_1}}
\end{figure*}

We study this model for different cases, corresponding to different parameter choices. In the first case we fix $-\Delta G^r(1)=-\Delta G^w(1)$ upon choosing $b_r=b_w=b$  and $c_r=c_w=1/20$. This enforces a kinetic discrimination regime \cite{Pigolotti2016} where the binding energy $-\Delta G^r(1)$ quantifies the degree of irreversibility. For highly irreversible processes, the error $\eta$ should mainly depend on the activation energy difference $\delta^r-\delta^r$, see Eq. \eqref{eq:integrals_saddle}. We also expect that the reaction speed should increase for more irreversible processes.  Equations \eqref{eq:eta_single_step} and \eqref{eq:v_definition} confirm such qualitative picture, see Figure \ref{fig:numerical_simulations_barrier_1}a and b.  Also in this case, { numerical simulations are in excellent quantitative agreement with our theory}, Fig. \ref{fig:numerical_simulations_barrier_1}c and \ref{fig:numerical_simulations_barrier_1}d.

\begin{figure*}
\includegraphics[width=\linewidth]{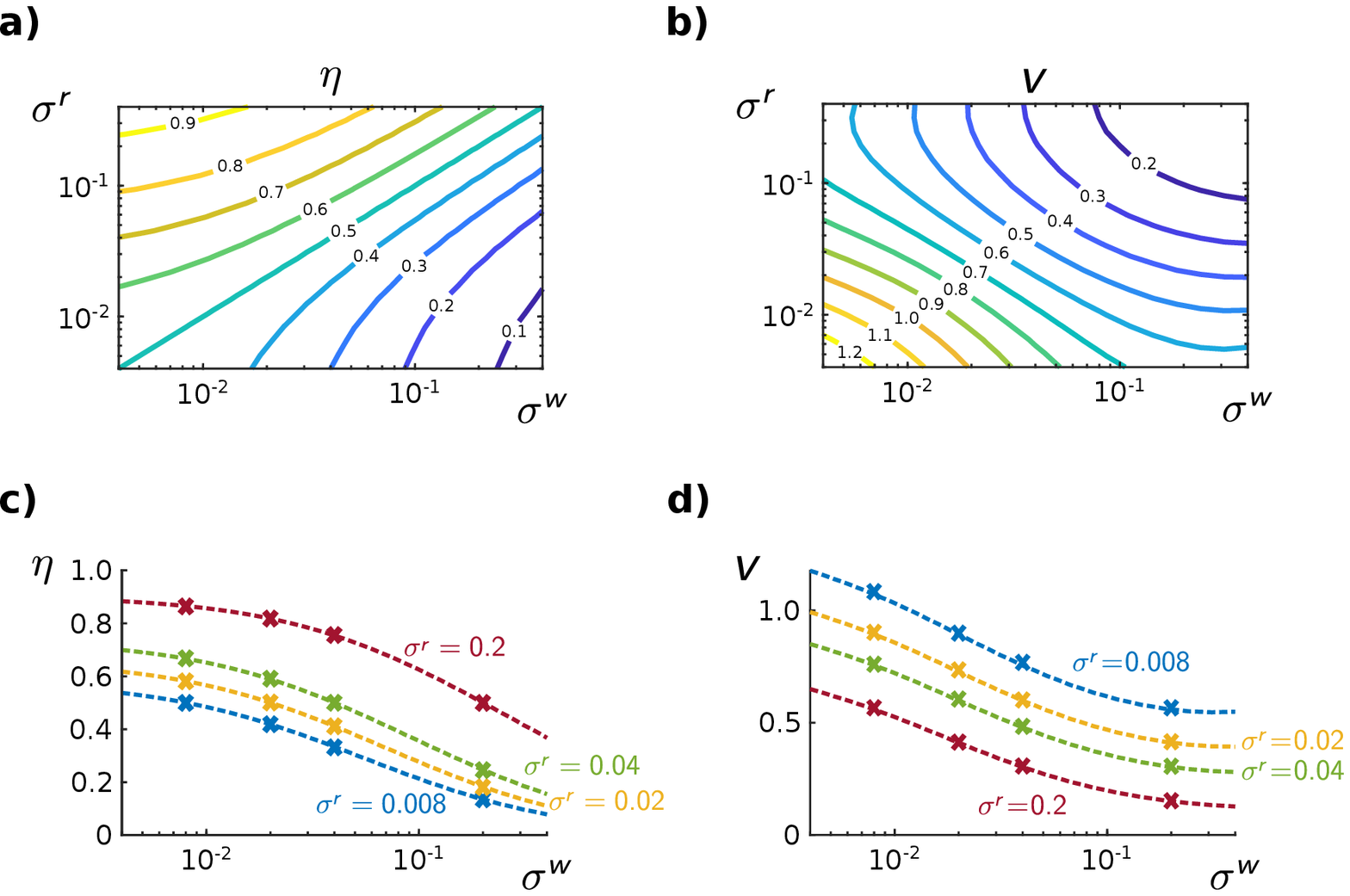}
\caption{Different barrier widths allow for kinetic discrimination in the absence of binding and activation energy differences. (Top) surface plots of $\eta$ and $v$ from Eqs.  \eqref{eq:eta_single_step} and \eqref{eq:v_definition} as a function of the barrier widths $\sigma_r$ and $\sigma_w$. To ensure that  $G^r(y)$ and $G^w(y)$ have approximately the same binding and activation energies we fixed $a_r=a_w=5$, $b_r=b_w=1$, and $c_r, c_w\leq 0.05$. (Bottom) Plots of $\eta$ and $v$ versus $\sigma_w$ for selected values of $\sigma^r$. Crosses corresponds to the average $\eta$ and $v$ values measured from $400$ simulations of the incorporation process with eq.\eqref{eq:Langevin_dynamics}. The Langevin dynamics was simulated with the weak 2.0 Runge-Kutta stochastic scheme \cite{kloeden2011numerical}. \label{fig:numerical_simulations_barrier_2}}
\end{figure*}

As a second  case, we fix $a_r=a_w=5$ and $b_r=b_w=1$. In this way we have that $-\Delta G^r(1)\approx -\Delta G^w(1)$ and $\delta^r\approx \delta^w$. Energetics alone would not permit monomer discrimination in this case \cite{Pigolotti2016}. However, Eq.  \eqref{eq:integrals_saddle} predicts that the difference in the barrier widths $\sigma_r$ and $\sigma_w$ should allow to discriminate $r$ and $w$ monomers (see Figure \ref{fig:numerical_simulations_barrier_2}.a). We confirmed the existence of such kinetic discrimination regime with numerical simulations, Fig. \ref{fig:numerical_simulations_barrier_2}.c. 

\section{Kinetic proofreading}\label{sec:kinetic_proofreading}

In this Section we sketch a generalization of our framework to include  kinetic proofreading \cite{Hopfield4135, NINIO1975587}. We assume that the reaction $h\rightleftharpoons hx$ can be decomposed into three sub-reactions
\begin{equation}\label{eq:proofreading_accomodation}
\schemestart
 $h$
 \arrow{<=>[$\scriptstyle p_{\rightarrow}^{1,x}$][$\scriptstyle p_{\leftarrow}^{1,x}$]}
 $hx^{*}$ \arrow(@c2--){<=>[*{0}$\scriptstyle p_{\leftarrow}^{3,x}$][*{0}$\scriptstyle p_{\rightarrow}^{3,x}$]}[-90,0.8] $h$ 
 \arrow(@c2--){<=>[$\scriptstyle p_{\rightarrow}^{2,x}$][$\scriptstyle p_{\leftarrow}^{2,x}$]} 
 $hx$
\schemestop
\end{equation}
where each sub-reaction occurs with  probabilities $p_{\rightarrow}^{i,x}$s and $p_{\leftarrow}^{i,x}$s, and $hx^*$ is an intermediate meta-stable complex. The extra pathway $hx^*\rightleftharpoons h$ represents kinetic proofreading. Such reaction can improve accuracy when driven towards the reactants $h$, so that wrong monomers undergo an additional checkpoint. \cite{Hopfield4135,Pigolotti2016}.

Every sub-reactions in Eq. \eqref{eq:proofreading_accomodation} is described by its own reaction coordinate $y$ which evolves according to a Langevin equation 
\begin{equation}\label{eq:Langevin_dynamics_general}
\frac{\mathrm{d}y}{\mathrm{d} t}=-\mu\frac{\mathrm{d}}{\mathrm{d} y} G^{i,x}(y) +\sqrt{2 D}~\xi(t)
\end{equation}
where $G^{i,hx}(y)$ is the free energy landscapes along the $i$-th sub-reaction. Also in this case we take $y\in[0,1]$ for all sub-reactions, with $y=1$ always in the direction of incorporation of monomer $x$. Similarly to Eq. \eqref{eq:free_energy_quasi_continuity}, we decompose the free energies for the sub-reactions as 
\begin{subequations}    
\begin{align}
    G^{1,h\tilde{x}x}(y)=&\Delta G^{1,x}(y)+G^{2,h\tilde{x}}(1)\\
    G^{2,h\tilde{x}x}(y)=&\Delta G^{2,x}(y)+G^{1,h\tilde{x}x}(1)\\
    G^{3,h\tilde{x}x}(y)=&\begin{cases}
                \Delta G^{3,x}(y)+G^{2,h\tilde{x}}(1)\quad \mbox{for  } hx^*\rightarrow h\\
                \Delta G^{3,x}(y)+G^{1,h\tilde{x}x}(1)\quad \mbox{for  } hx^*\leftarrow h
                \end{cases}
\end{align}
\end{subequations}
where we specified that monomer $\tilde{x}$ was incorporated before attempting to incorporate monomer $x$. Here, $G^{3,h\tilde{x}x}(y)$ depend on the direction of the sub-reaction because the heteropolymer total energy now depends also on the sequence of sub-reactions.

We now compute the probabilities $p_{\rightarrow}^{i,x}$s and $p_{\leftarrow}^{i,x}$ with $i\in\{1,2,3\}$ from Eq. \eqref{eq:Langevin_dynamics_general} with the same procedure which leads to  Eq.\eqref{eq:p_reaction_coordinate}. { This yields }
\begin{subequations}\label{eq:kinetic_rates_Hopfield}
\begin{align}
p_\rightarrow^{i,x}=&\epsilon\,\pi_{\epsilon,0}^{i,x}\,\frac{1}{\int_{0}^{1}\exp[\frac{\Delta G^{i,x}(y)}{\kt}]\,\mathrm{d}y}+\mathcal{O}(\epsilon)\\
p_\leftarrow^{i,x}=&\epsilon\,\pi_{1-\epsilon,1}^{i,x} \,\frac{e^{\frac{\Delta G^{i,x}(1)}{\kt}}}{\int_{0}^{1}\exp[\frac{\Delta G^{i,x}(y)}{\kt}]\,\mathrm{d}y}+\mathcal{O}(\epsilon).
\end{align}
\end{subequations}
with
\begin{subequations}\label{eq:split_probabilities_complex}
\begin{align}
\pi_{\epsilon,0}^{1,x}=\pi_{1-\epsilon,1}^{2,x}=\pi_{\epsilon,0}^{3,x}&=\frac{1}{5}\\
p_{1-\epsilon,1}^{1,x}=\pi_{\epsilon,0}^{2,x}=\pi_{\epsilon,0}^{3,x}&=\frac{1}{3}.
\end{align}
\end{subequations}
Equations \eqref{eq:split_probabilities_complex} state that the all sub-reactions from reactants $h$ and $hx^*$ respectively can start with equal probabilities.

To obtain an equation for $\eta$, we need to compute the effective incorporation and removal probabilities $p_\rightarrow^x$ and $p_\leftarrow^x$ in Eq. \eqref{eq:eta_equation} from Eqs. \eqref{eq:kinetic_rates_Hopfield} and \eqref{eq:split_probabilities_complex}. To this end we { assume that the reactions in Eq.\eqref{eq:proofreading_accomodation} are at steady state. We then use adiabatic elimination \cite{doi:10.1063/1.2907242} to obtain (see Appendix \ref{sec:adiab_elim_kinetic_proof})}
\begin{subequations}\label{eq:effective_rates_Hopfield}
\begin{align}
p_{\rightarrow}^x=&\frac{p_{\rightarrow}^{2,x} (p_{\rightarrow}^{1,x} + p_{\rightarrow}^{3,x})}{p_{\leftarrow}^{1,x} + p_{\leftarrow}^{3,x} + p_{\rightarrow}^{2,x}}\\
p_{\leftarrow}^x=&\frac{p_{\leftarrow}^{2,x} (p_{\leftarrow}^{1,x} + p_{\leftarrow}^{3,x})}{p_{\leftarrow}^{1,x} + p_{\leftarrow}^{3,x} + p_{\rightarrow}^{2,x}}
\end{align}
\end{subequations}
{ Substituting Eqs. \eqref{eq:kinetic_rates_Hopfield} and \eqref{eq:effective_rates_Hopfield} in Eq. \eqref{eq:eta_equation} finally provides an expression for $\eta$ in terms of the free energy landscapes $G^{i,x}(y)$.}

\section{Conclusions}\label{sec:conclusions}

In this paper, we described assembly of heteropolymers by means of continuous reaction coordinates. In the simplest cases, our results are consistent with those derived for reactions occurring in discrete steps
\cite{andrieux2008nonequilibrium, doi:10.1063/1.4890821, Banerjee5183, PhysRevX.5.041039, PhysRevLett.110.188101, murugan2012speed, Pigolotti2016, bennett1979dissipation}. Moreover, our formalism reveals discrimination mechanisms that are not easily described with discrete reactions. One example is the possibility to discriminate according to barrier widths, as described by Eq. \eqref{eq:integrals_saddle} and confirmed in simulations, Fig. \ref{fig:numerical_simulations_barrier_2}c. 

{ For simplicity, in this paper we developed our formalism by means of a  reaction coordinate characterized by a Markovian dynamic.  In general, only specific projection techniques yield reaction coordinates with negligible non-Markovian contributions \cite{doi:10.1002/wcms.1276, doi:10.1063/1.4890367, doi:10.1021/jp060039b, Krivov2018}, and the resulting Langevin equation might not be in the form of Eq.  \eqref{eq:Langevin_dynamics}}. { Our framework can be adapted to such situations as well as to non-Markovian reaction coordinates, describing for example enzymes undergoing slow conformational changes. }

The framework described here is microscopically reversible. This allows to characterize non-equilibrium work and heat exchanges during the polymerization process  from the diffusive dynamics of the reaction coordinate, similarly to recent studies of the ATP synthase \cite{PhysRevE.99.012119, Kasper2019} and small-scale technological devices \cite{Neri_2015,  López-Suárez2016}.  This analysis would permit to characterize thermodynamic limits of information processing of these processes \cite{Berut2012, PhysRevLett.113.190601, Chiuchi__2015,PhysRevX.5.041039}.

\begin{acknowledgments}
This work was supported by JSPS KAKENHI Grant Number JP18K03473 (to DC and SP).
\end{acknowledgments}

\appendix   

{\section{First passage time probabilities at the nodes}\label{sec:nodes}

Because of detailed balance, the probabilities $\pi_{\epsilon,0}^x$ and $\pi_{1-\epsilon,1}^x$ are related to the free energy difference when passing from one edge of the reaction network to another, i.e. 
\begin{eqnarray}\label{eq:energy_dependence_probabilities}
\pi_{\epsilon,0}^x&\propto& \exp\left[-\frac{G^{h\tilde{x}x}(\epsilon)}{\kt}\right]\nonumber\\
\pi_{1-\epsilon,1}^{\tilde{x}}&\propto& \exp\left[-\frac{G^{h\tilde{x}}(1-\epsilon)}{\kt}\right] .
\end{eqnarray}
where we specified that the monomer $\tilde{x}\in\{r,w\}$ was incorporated before monomer $x\in \{r,w\}$. After the incorporation of $\tilde{x}$, the enzyme can catalyze three reactions: removal of $\tilde{x}$ or incorporation of either $r$ or $w$. The probabilities of these three events must be normalized
\begin{equation}\label{eq:normalization_condition}
\pi_{1-\epsilon,1}^{\tilde{x}}+\pi_{\epsilon,0}^r+\pi_{\epsilon,0}^w=1.
\end{equation}
Combining Eqs.\eqref{eq:energy_dependence_probabilities}-\eqref{eq:normalization_condition} gives
\begin{subequations}\label{eq:proto_pis_initiation energy}
\begin{align}
\pi_{\epsilon,0}^x=&  \frac{\exp\left[-\frac{ G^{h\tilde{x}x}(\epsilon)-G^{h\tilde{x}}(1-\epsilon)}{\kt}\right]}{1+\exp\left[-\frac{G^{h\tilde{x}r}(\epsilon)-G^{h\tilde{x}}(1-\epsilon)}{\kt}\right]+\exp\left[-\frac{G^{h\tilde{x}w}(\epsilon)-G^{h\tilde{x}}(1-\epsilon)}{\kt}\right]}\\
\pi_{1-\epsilon,1}^{\tilde{x}}=& \frac{1}{1+\exp\left[-\frac{ G^{h\tilde{x}r}(\epsilon)-G^{\tilde{x}}(1-\epsilon)}{\kt}\right]+\exp\left[-\frac{G^{h\tilde{x}w}(\epsilon)-G^{h\tilde{x}}(1-\epsilon)}{\kt}\right]}.
\end{align}
\end{subequations}
Substituting Eq.\eqref{eq:free_energy_quasi_continuity} into Eq.\eqref{eq:proto_pis_initiation energy}, taking the limit of small $\epsilon$, using the continuity of $G^{h\tilde{x}x}(y)$ and then renaming $\tilde{x}$ with $x$ finally gives Eq.\eqref{eq:pis_initiation energy}.

\section{Effective incorporation and removal probabilities}
\label{sec:effective_probabilities}

The dynamics of the reaction coordinate $y$ in the coarse grained description of Eq.\eqref{eq:four_state_dynamics} obeys a Markov chain
\begin{subequations}\label{eq:ME_reaction_coordinate}
\begin{align}
P_0^{x}(\nu+1)=&\pi_{0,\epsilon}^{x}P_\epsilon^{x}(\nu)+\left[1-\pi_{\epsilon,0}^{x}\right]P_0^{x}(\nu)\label{eq:ME_1}+\mbox{external fluxes }\\
\begin{split}
P_\epsilon^{x}(\nu+1)=&\pi_{\epsilon,0}^{x}P_0^{x}(\nu)+\pi_{\epsilon,1-\epsilon}^{x}P_{1-\epsilon}^{x}(\nu)+\left[1-\left(\pi_{0,\epsilon}^{x}+\pi_{1-\epsilon,\epsilon}^{x}\right)\right]P_\epsilon^{x}(\nu)\label{eq:ME_2}
\end{split}\\
\begin{split}
P_{1-\epsilon}^{x}(\nu+1)=&\pi_{1-\epsilon,\epsilon}^{x}P_\epsilon^{x}(\nu)+\pi_{1-\epsilon,1}^{x}P_{1}^{x}(\nu)+\left[1-\left(\pi_{\epsilon,1-\epsilon}^{x}+\pi_{1,1-\epsilon}^{x}\right)\right]P_{1-\epsilon}^{x}(\nu)\label{eq:ME_3}
\end{split}\\
P_1^{x}(\nu+1)=&\pi_{1,1-\epsilon}^{x}P_{1-\epsilon}^{x}(\nu)+\left[1-\pi_{1-\epsilon,1}^{x}\right]P_1^{x}(\nu)+\mbox{external fluxes}.\label{eq:ME_4}
\end{align}
\end{subequations}
where the first-passage probabilities appear as transition probabilities, and the quantities $P^x_0(\nu)$, $P_\epsilon(\nu)$, $P_{1-\epsilon}(\nu)$, and $P_1(\nu)$ are the probabilities that the reaction coordinate reaches the point $y=0$, $y=\epsilon$, $y=1-\epsilon$ and $y=1$ after $\nu$ consecutive transitions respectively. The \emph{external fluxes} in Eqs. \eqref{eq:ME_1} and \eqref{eq:ME_4} are the probability fluxes from the remaining reactions which originate from the nodes $y=0$ and $y=1$ in the network of Figure \ref{fig:tree}.a. 

To simplify Eq. \eqref{eq:ME_reaction_coordinate} we perform adiabatic elimination \cite{doi:10.1063/1.2907242} of the intermediate states $y=\epsilon$ and $y=1-\epsilon$: we impose the steady state regime $P_\epsilon^x(\nu+1)=P_\epsilon^x(\nu)$ and $P_\epsilon^x(\nu+1)=P_\epsilon^x(\nu)$ in Eqs.\eqref{eq:ME_2} and \eqref{eq:ME_3} respectively, we solve  Eqs.\eqref{eq:ME_2}-\eqref{eq:ME_3} for $P_\epsilon^{x}(\nu)$ and $P_{1-\epsilon}^{x}(\nu)$, and we finally substitute the result back into Eqs.\eqref{eq:ME_1}, \eqref{eq:ME_4}. This yield the effective Markov chain
\begin{subequations}\label{eq:ME_reaction_coordinate_adiabatic}
\begin{align}
P_0^{x}(\nu+1)=&p_{\leftarrow}^{x}P_1^{x}(\nu)+\left[1-p_{\rightarrow}^{x}\right]P_0^{x}(\nu)+\mbox{external fluxes}\\
P_1^{x}(\nu+1)=&p_{\rightarrow}^{x}P_0^{x}(\nu)+\left[1-p_{\leftarrow}^{x}\right]P_1^{x}(\nu)+\mbox{external fluxes }.
\end{align}
\end{subequations}
where we have defined the effective probabilities $p_\rightarrow^{x}$ and $p_\leftarrow^{x}$ to incorporate a monomer ($h\rightarrow hx$) and remove a monomer ($h\leftarrow hx$) respectively as \begin{subequations}\label{eq:single_reactioN^rate}
\begin{align}
p_\rightarrow^{x}=&\frac{\pi_{1,1-\epsilon}^{x}\pi_{1-\epsilon,\epsilon}^{x}\pi_{\epsilon,0}^{x}}{\pi_{1,1-\epsilon}^{x}\ \pi_{1-\epsilon,\epsilon}^{x}+\pi_{0,\epsilon}^{x}\ \pi_{\epsilon,1-\epsilon}^{x}+\pi_{0,\epsilon}^{x}\ \pi_{1,1-\epsilon}^{x}}\\
p_\leftarrow^{x}=&\frac{\pi_{0,\epsilon}^{x}\ \pi_{\epsilon,1-\epsilon}^{x}\ \pi_{1-\epsilon,1}^{x}}{\pi_{1,1-\epsilon}^{x}\ \pi_{1-\epsilon,\epsilon}^{x}+\pi_{0,\epsilon}^{x}\ \pi_{\epsilon,1-\epsilon}^{x}+\pi_{0,\epsilon}^{x}\ \pi_{1,1-\epsilon}^{x}}.
\end{align}
\end{subequations}
Substituting Eq.\eqref{eq:pis_initiation energy}-\eqref{eq:psi} into Eq. \eqref{eq:single_reactioN^rate} and then expanding for small $\epsilon$ finally gives Eq. \eqref{eq:p_reaction_coordinate}.}

\section{Derivation of the polymerization speed via reaction coordinates.}
\label{sec:derive_velocity}
\begin{figure}
\includegraphics[width=\linewidth]{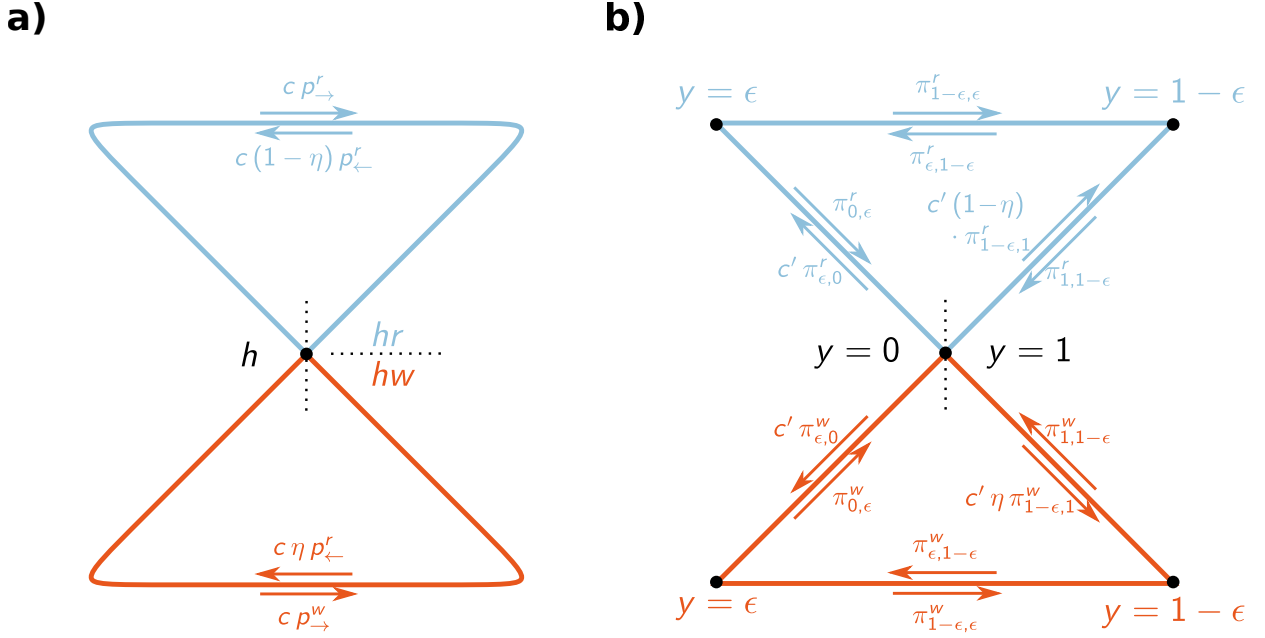}
\caption{Mean field representation of the polymerization process. a) Mean field version of the heteropolymer assembly in Figure \ref{fig:tree}.a where the enzyme can remove any monomer in $h$. Removal involves $r$ and $w$ monomers $(1-\eta)$ and $\eta$ times respectively. The probabilities $p^x_\rightarrow$, $p^x_\leftarrow$ are defined as in Eq. \eqref{eq:p_reaction_coordinate}. The constant $c=[p_\rightarrow^r+(1-\eta) p_\leftarrow^r+p_\rightarrow^w+\eta p_\leftarrow^w]^{-1}$  normalizes the probabilities and is defined as in Eqs. \eqref{eq:count_reactions}. b) Same as in panel a, but we have now explicitly introduced the intermediate reaction coordinate values $y=\epsilon$ and $y=1-\epsilon$, as well as the transition probabilities $\pi_{y,\tilde{y}}^x$ defined as in Eqs. \eqref{eq:pis_initiation energy} and \eqref{eq:pis_first_passage}. The constant $c'=[\pi_{\epsilon,0}^r+(1-\eta)\pi_{1-\epsilon,1}^r+\pi_{\epsilon,0}^w+\eta \pi_{1-\epsilon,1}^w]$  normalizes the probabilities exiting from the central node. \label{fig:meanfield}}
\end{figure}

To derive the polymerization speed, we consider a mean field formulation of the polymerization process in Figure \ref{fig:tree}.a where the enzyme can remove any monomer in the copy heteropolymer. Removal of $r$ and $w$ monomers occurs with probabilities $1-\eta$ and $\eta$ respectively. This assumption simplifies the reaction tree of Figure \ref{fig:tree}.a into the closed network of Fig. \ref{fig:meanfield}.a,
where the incorporation and removal probabilities $p_\rightarrow^x$ and $p_\leftarrow^x$ are defined as in Eq. \eqref{eq:p_reaction_coordinate}.

We now introduce the reaction coordinate in this mean field description, Fig. \ref{fig:meanfield}.b. For later convenience, we also consider the 
values of the reaction coordinate $y=0$ $y=\epsilon$, $y=1-\epsilon$ and $y=1$ together with the probabilities  $\pi_{\tilde{y},y}^{x}$s defined in Eqs.\eqref{eq:pis_initiation energy} and \eqref{eq:pis_first_passage}.

Using the scheme in Figure \ref{fig:meanfield}.b, we define the probability $P_{0,1}(\zeta)$ that $y=0$ or $y=1$ after $\zeta$ consecutive transitions, and the probabilities $P_\epsilon^r(\zeta)$, $P_{1-\epsilon}^r(\zeta)$, $P_\epsilon^w(\zeta)$, $P_{1-\epsilon}^w(\zeta)$ that $y=\epsilon$ or $y=1-\epsilon$ for the $r$ and $w$ monomer after $\zeta$ consecutive transitions. These probabilities evolves according to the Markov chain
\begin{equation}\label{eq:ME_reaction_coordinate_mean_field}
\vec{P}(\zeta+1)=\mathtt{A}\,\vec{P}(\zeta)
\end{equation}
where 
\begin{equation}
\vec{P}(\zeta)=\left(P_{0,1}(\zeta), P_\epsilon^r(\zeta), P_{1-\epsilon}^r(\zeta), P_\epsilon^w(\zeta), P_{1-\epsilon}^w(\zeta)\right)^\mathtt{T}
\end{equation}
and 
\begin{equation}\label{eq:A_matrix}
\mathtt{A}=\begin{bmatrix}
0 & \pi_{0,\epsilon}^r & \pi_{1,1-\epsilon}^r & \pi_{0,\epsilon}^w & \pi_{1,1-\epsilon}^w\\
c'\,\pi_{\epsilon,0}^r & 0 & \pi_{\epsilon,1-\epsilon}^r & 0 & 0\\
c'\,(1-\eta)\pi_{1-\epsilon,1}^r & \pi_{1-\epsilon,\epsilon}^r & 0 & 0 & 0\\
c'\,\pi_{\epsilon,0}^w & 0 & 0 & 0 & \pi_{\epsilon,1-\epsilon}^w\\
c'\,\eta \pi_{1-\epsilon,1}^w & 0 & 0 & \pi_{1-\epsilon,\epsilon}^w & 0 
\end{bmatrix}
\end{equation}
where $c'=[\pi_{\epsilon,0}^r+(1-\eta)\pi_{1-\epsilon,1}^r+\pi_{\epsilon,0}^w+\eta \pi_{1-\epsilon,1}^w]$ is a normalization constant. We now define the matrices
\begin{subequations}\label{eq:increment_matrices}
\begin{align}
\mathtt{J}^N&=\frac{1}{3}\begin{bmatrix}
0 & -1 & +1 & -1 & +1\\
+1 & 0 & -1 & 0 & 0\\
-1 & +1 & 0 & 0 & 0\\
+1 & 0 & 0 & 0 & -1\\
-1 & 0 & 0 & +1 & 0 
\end{bmatrix}\\ 
\mathtt{J}^{\mathcal{T}}&=\begin{bmatrix}
0 & \langle\mathrm{d}\tau\rangle_{0,\epsilon}^r & \langle\mathrm{d}\tau\rangle_{1,1-\epsilon}^r & \langle\mathrm{d}\tau\rangle_{0,\epsilon}^w & \langle\mathrm{d}\tau\rangle_{1,1-\epsilon}^w\\
\langle\mathrm{d}\tau\rangle_{\epsilon,0}^r & 0 & \langle\mathrm{d}\tau\rangle_{\epsilon,1-\epsilon}^r & 0 & 0\\
\langle\mathrm{d}\tau\rangle_{1-\epsilon,1}^r & \langle\mathrm{d}\tau\rangle_{1-\epsilon,\epsilon}^r & 0 & 0 & 0\\
\langle\mathrm{d}\tau\rangle_{\epsilon,0}^w & 0 & 0 & 0 & \langle\mathrm{d}\tau\rangle_{\epsilon,1-\epsilon}^w\\
\langle\mathrm{d}\tau\rangle_{1-\epsilon,1}^w & 0 & 0 & \langle\mathrm{d}\tau\rangle_{1-\epsilon,\epsilon} & 0 
\end{bmatrix}
\end{align}
\end{subequations}
which contain the contribution of each transition to the heteropolymer length $N$ and the polymerization time $\mathcal{T}$. The time increments $\langle\mathrm{d}\tau\rangle_{y,\tilde{y}}^x$ in $\mathtt{J}^{\mathcal{T}}$ are the first passage times from $\tilde{y}$ to $y$ \cite{gardiner2009stochastic}. In particular we have that
\begin{subequations}
\begin{align}
\langle\tau\rangle_{0,\epsilon}^x=&\frac{1}{D}\ \left(\Phi_\rightarrow^x(0)-\frac{\int_0^1 \Phi_\rightarrow^x(y) \mathrm{d}y }{\int_0^1  \psi^x(y)\mathrm{d}y}\right)\epsilon+\mathcal{O}\left(\epsilon^2\right)\\
\langle\tau\rangle_{1-\epsilon,\epsilon}=&\frac{1}{D}\  \int_0^1 \Phi_\leftarrow^x(y) \mathrm{d}y +\mathcal{O}\left(\epsilon	\right)\\
\langle\tau\rangle_{\epsilon,1-\epsilon}=&\frac{1}{D}\  \int_0^1 \Phi_\rightarrow^x(y) \mathrm{d}y +\mathcal{O}\left(\epsilon	\right)\\
\langle\tau\rangle_{1,1-\epsilon}^x=&\frac{1}{D}\ \left(\Phi_\leftarrow^x(1)-\left(e^{\frac{\Delta G^x(1)}{\kt}}\right)\frac{\int_0^1 \Phi_\leftarrow^x(y) \mathrm{d}y }{\int_0^1  \psi^x(y)\mathrm{d}y}\right)\epsilon+\mathcal{O}\left(\epsilon^2\right)
\end{align}
\end{subequations}
with
\begin{subequations}
\begin{align}
    \Phi_\rightarrow^x(y)&=\frac{\psi^x(y)\int_{y}^{1}\int_{u}^1\frac{\psi^x(z)}{\psi^x(u)}\mathrm{d}z\mathrm{d}u   }{\int_0^1 \psi^x(y)\mathrm{d}y}\\
    \Phi_\leftarrow^x(y)&=\frac{\psi^x(y)\int_{0}^{y}\int_{0}^u\frac{\psi^x(z)}{\psi^x(u)}\mathrm{d}z\mathrm{d}u   }{\int_0^1 \psi^x(y)\mathrm{d}y}.
\end{align}
\end{subequations}
{ The remaining first passage times $\langle\mathrm{d}\tau\rangle_{\epsilon,0}^r$, $\langle\mathrm{d}\tau\rangle_{1-\epsilon,1}^r$, $\langle\mathrm{d}\tau\rangle_{\epsilon,0}^w$ and $\langle\mathrm{d}\tau\rangle_{1-\epsilon,1}^w$, are assumed equal to zero for simplicity. Physically, this assumption is justified when binding and unbinding of monomers is much faster than processing a monomer into a finalized incorporation.}

Using Eq.\eqref{eq:increment_matrices} we define the tilted matrix $\mathtt{B}$ with components
\begin{equation}\label{eq:tilted_matrix}
\mathtt{B}_{i,j}=\mathtt{A}_{i,j}\exp\left[q_N \mathtt{J}^N_{i,j}+q_\tau \mathtt{J}^{\tau}_{i,j} \right]
\end{equation}
and dummy variables $q_N$, and $q_\tau$. For large values of $\zeta$, the largest eigenvalue of $\mathtt{B}$ coincides with the scaled cumulant generating function of $N$ and $\mathcal{T}$, see \cite{TOUCHETTE20091}. The implicit function theorem then implies  
\begin{subequations}
\begin{align}
N \approx -\zeta\left.\frac{\partial_{q_{N}}\det\left[\mathtt{B}-\lambda\mathtt{I}\right]}{\partial_{\lambda}\det\left[\mathtt{B}-\lambda\mathtt{I}\right]}\right|_{q_N=q_\tau=0, \lambda=1}\\
\mathcal{T} \approx -\zeta\left.\frac{\partial_{q_{\tau}}\det\left[\mathtt{B}-\lambda\mathtt{I}\right]}{\partial_{\lambda}\det\left[\mathtt{B}-\lambda\mathtt{I}\right]}\right|_{q_N=q_\tau=0, \lambda=1}
\end{align}
\end{subequations}
where $\det\left[\mathtt{B}-\lambda\mathtt{I}\right]$ is the characteristic polynomial of $\mathtt{B}$. To compute $v$ we finally use that 
\begin{equation}\label{eq:v_equation_large_deviation}
v=\frac{N}{\mathcal{T}}=\left.\frac{\partial_{q_N}\det\left[\mathtt{B}-\lambda\mathtt{I}\right]}{\partial_{q_\tau}\det\left[\mathtt{B}-\lambda\mathtt{I}\right]}\right|_{q_N=q_\tau=0, \lambda=1}.
\end{equation}
which is equivalent to Eq.\eqref{eq:proto_v_definition}. Substituting Eqs.\eqref{eq:pis_initiation energy}, \eqref{eq:pis_first_passage}, \eqref{eq:A_matrix}, \eqref{eq:increment_matrices} and \eqref{eq:tilted_matrix} into Eqs.\eqref{eq:v_equation_large_deviation} and then taking the leading order for small $\epsilon$  yields Eq. \eqref{eq:v_definition}, where
\begin{equation}\label{eq:tau}
\begin{aligned}
\frac{\langle \tau\rangle D}{c}=& \left(p_\rightarrow^r -(1-\eta) p_\leftarrow^r\right)\left(\int_{0}^1\left[\Phi_\leftarrow^r(y)-\Phi_\rightarrow^r(y)\right]\mathrm{d}y\right)+ \left(p_\rightarrow^w -\eta p_\leftarrow^w\right)\left(\int_{0}^1\left[\Phi_\leftarrow^w(y)-\Phi_\rightarrow^w(y)\right]\mathrm{d}y\right)
\\&{+\frac{\epsilon}{3}\,\Bigg( \Phi_\rightarrow^r(0)+\Phi_\rightarrow^w(0)+(1-\eta)\Phi^r_\leftarrow(1)+\eta\Phi^w_\leftarrow(1)\Bigg) + \mathcal{O}\left(\epsilon^2\right)}
\end{aligned}
\end{equation}
and $c$ is defined as in Eq. \eqref{eq:count_reactions}.

{ \section{Effetive incorporation and removal probabilities for the Kinetic proofreading example}
\label{sec:adiab_elim_kinetic_proof}
To compute the incorporation and removal probabilities for the Kinetic proofreading case we mimic the procedure that leads to Eq.\eqref{eq:p_reaction_coordinate}. We consider the probabilities $P_{h}(\xi)$, $P_{hx^*}(\xi)$ and $P_{hx}(\xi)$ to obtain the reactants $h$, $hx^*$, and $hx$ after $\xi$ sub-reactions of Eq. \eqref{eq:proofreading_accomodation}. These probabilities evolve according to the Markov chain
\begin{subequations}\label{eq:ME_proofreading/accomodation}
\begin{align}
P_{h}(\xi+1)&=p_{\leftarrow}^{1,x}P_{hx^x}(\xi)+\left[1-\left(p_\rightarrow^{1,x}+p_\rightarrow^{3,x}\right)\right]P_{h}(\xi)\label{eq:ME_E+h+x}+\mbox{external fluxes}\\
P_{hx^*}(\xi+1)&=p_{\rightarrow}^{1,x}P_{h}(\xi)+p_{\leftarrow}^{2,x}P_{hx}(\xi)+p_{\rightarrow}^{3,x}P_{h}(\xi)+\left[1-\left(p_\leftarrow^{1,x}+p_\rightarrow^{2,x}+p_\leftarrow^{3,x}\right) \right]P_{hx^*}(\xi)\label{eq:ME_Ehx}\\
P_{hx}(\xi+1)&=p_{\rightarrow}^{2,x}P_{hx^{*}}(\xi)+\left[1-p_\leftarrow^{2,x}\right]P_{hx}(\xi)+\mbox{external fluxes}\label{eq:ME_E+hx},
\end{align}
\end{subequations}
where the \emph{external fluxes} are the probability fluxes of the other sub-reactions entering the nodes $y=0$ and $y=1$. At steady state, we simplify Eq. \eqref{eq:ME_proofreading/accomodation} with adiabatic elimination\cite{doi:10.1063/1.2907242}:  we impose $P_{hx^*}(\xi+1)=P_{hx^*}(\xi)$ into Eq. \eqref{eq:ME_Ehx}, solve it for $P_{hx^*}(\xi)$ and substitute the solution in Eqs. \eqref{eq:ME_E+h+x} and \eqref{eq:ME_E+hx}. This yields, after some rearrangements, 
\begin{subequations}
\begin{align}
P_{h+x}(\xi+1)&=p_{\leftarrow}^xP_{hx}(\xi)+\left[1-p_{\rightarrow}^x\right]P_{h+x}(\xi)\\
P_{hx}(\xi+1)&=p_{\rightarrow}^xP_{h+x}(\xi)+\left[1-p_{\leftarrow}^x\right]P_{hx}(\xi)
\end{align}
\end{subequations}
with effective incorporation/removal probabilities $p_\rightarrow^x$ and $p_\leftarrow^x$ defined as in Eq.\eqref{eq:effective_rates_Hopfield}.}

\nocite{*}
\bibliography{references}

\end{document}